\documentclass[twocolumn,prl,aps,showpacs,floatfix]{revtex4}

\usepackage{amsfonts,amsmath,graphicx,latexsym}
\usepackage[utf8]{inputenc}
\usepackage{epsfig}
\usepackage{psfrag}

\newcommand{\vc}[1]{\mathbf{#1}}

\begin{document}
\title{Gap and screening in Raman scattering of a 
Bose condensed gas}
\author{Patrick Navez$^1$, Kai Bongs$^2$}
\affiliation{$^1$
Universitaet Duisburg-Essen, Lotharstrasse 1, 47057 Duisburg, Germany\\
$^2$ University of Birmingham, Edgbaston, Birmingham, B15 2TT, England}

\begin{abstract}
We propose different spectroscopic methods to explore the nature of the thermal excitations 
of a trapped Bose condensed gas: 1) a four photon process to probe the uniform 
region in the trap center: 2) a stimulated Raman process in order 
to analyze the influence of a momentum transfer in the resulting 
scattered atom momentum distribution.   
We apply these methods to address specifically the energy spectrum and the scattering 
amplitude of these excitations in a transition between two 
hyperfine levels of the gas atoms. 
In particular, we exemplify the potential offered by these 
proposed techniques by contrasting the spectrum expected,
from the {\it non conserving} Bogoliubov approximation valid for weak depletion, 
to the spectrum of the finite temperature extensions like  
the {\it conserving} generalized random phase approximation 
(GRPA). Both predict the existence of 
the Bogoliubov collective excitations but 
the GRPA approximation distinguishes them from the single atom excitations with 
a gapped and parabolic dispersion relation and accounts for
the dynamical screening of any external perturbation applied to the gas. 
We propose two feasible experiments, one concerns the observation of the 
gap associated to this second branch of excitations and the other deals with 
this screening effect.
\end{abstract}

\maketitle


\section{Introduction}
\subsection{Nature of the elementary excitations}
The experimental discovery of the condensation of a Bose gas  
has confirmed the existence of the 
phonon-like nature of the collective excitations
\cite{Ketterle, Stringari}. The obtained measured energy spectrum not only 
is gapless as stated from the Hugenholtz-Pines theorem but is also 
in perfect agreement with the prediction of 
the Bogoliubov approach at zero temperature \cite{HP}. 
However, a second fundamental 
question arises as to whether these collective excitations are the elementary 
building constituents for the normal part of the fluid as assumed 
in the Bogoliubov approximation. Most 
standard textbooks rely on this quasiparticle hypothesis in order to 
determine the finite temperature gas properties \cite{Leggett, books}.
In contrast, in the theoretical description of a 
plasma, distinction is made between the elementary 
excitations (ions) and the collective 
ones (plasmons). As discussed in previous works 
\cite{gap,gap2,condenson,Graham}, there are no 
fundamental reasons to exclude this 
distinction also in a Bose gas.

Precisely, suppose a bulk gas of total and condensed densities $n$ and $n_0$ 
embedded in an volume $V$ where atoms of mass $m$ interact through the s wave channel 
with a scattering length $a$. The Bogoliubov approximation predicts 
that the elementary excitations of momentum $\vc{k}$ are phonon-like with a dispersion relation 
given by $\epsilon_{1,\vc{k}}^B=\sqrt{2gn_\vc{0}\epsilon^{}_{\vc{k}}+ \epsilon^{2}_{\vc{k}}}$ 
where $\epsilon_\vc{k}=\hbar^2\vc{k}^2/2m$ and $g=4\pi a \hbar^2/m$. 
Nevertheless, its {\it non conserving} property (violation of the mass conservation law) 
\cite{HM} restricts 
its validity for a weakly depleted Bose gas and thus limits its use to low temperature.
As opposed to that, 
the so-called generalized random phase approximation 
(GRPA) or equivalently the time-dependent Hartree-Fock (TDHF) approximation is instead 
{\it conserving} and valid for the whole range of temperature. 
This alternative approach distinguishes explicitly these collective phonon-like excitations  
from the atom-like elementary excitations 
with the parabolic dispersion relation $\epsilon^{HF}_{1,\vc{k}}=\epsilon_\vc{k}
+g(2n- n_\vc{0}\delta_{\vc{k},\vc{0}})$ \cite{Reidl,Levitov,Zhang,gap2}.
The constant term corresponds to the Hartree and Fock (HF) mean field energy part and 
takes into account the absence of exchange interaction 
energy between condensed atoms. Therefore, an energy gap exists between the 
thermal and condensed atoms  
$\epsilon^{HF}_{1,\vc{k}}-\epsilon^{HF}_{1,\vc{0}}= gn_\vc{0}+\epsilon_\vc{k}$. 

\subsection{Superfluidity due to total screening}

Another important reason to discriminate among the various 
theoretical approaches is to have an improved 
understanding of the superfluidity phenomenon. More precisely, we would like 
to answer the following question: Why, from a kinetic point of view, a superfluid 
can remain in a metastable motion without converting its kinetic energy into heat? 
Many explanations have been provided but, according to \cite{Leggett}, 
{\it the situation is not entirely clear} as far as kinetic theory is concerned. 

Instead, the equilibrium aspects based on the 
ensemble approach of the superfluid phenomenon of 
a Bose condensed gas are well understood.  
Using the $\eta$ ensemble which breaks the $U(1)$ symmetry associated 
to the particle number conservation, 
one can describe the superfluid motion (condensed mode)
relatively to the normal fluid (non condensed modes) \cite{Graham}. 
Such a relative motion should not be considered as an equilibrium state 
but as a metastable state possibly subject to relaxation of a 
state of lower energy. Unfortunately, the ensemble approach  
does not explain the physical reasons for such a metastability.
It just tells that an artificial breaking of symmetry allows you such a 
description. Only a non-equilibrium treatment can provide these explanations 
and therefore confirm 
the validity of the assumptions used in the ensemble approach. 

The kinetic theory so far developed in the Bogoliubov approximation 
allows for such a metastability in the weak depletion limit \cite{Kirkpatrick}.
Particle exchange between the normal and superfluid 
are regulated through a balanced Beliaev process 
of transforming one collective 
excitations to two collective excitations.
A complete different 
scenario appears in the GRPA as it 
accounts for the dynamical screening of  
any external time-dependent potential that perturbs the gas atoms
\cite{condenson,gap2}. The ability of the macroscopic condensed wave function 
to deform locally its profile allows for a screening of any external perturbation 
that affects the energy transition probability of any 
atom-like excitation. In particular, under some stability conditions 
\cite{condenson}, a total screening 
forbids individual energy transitions involving a condensed atom. 
In this sense, the condensed atoms are {\it gregarious} since they 
respond only collectively to a perturbation via the creation of a phonon-like excitation. 
If the external potential originates from the presence of 
another thermal atom, this total screening prevents 
the binary collision between this thermal atom and any condensed 
one. Therefore, contrary to the Bogoliubov approach, the metastability of the
relative motion between the normal and super fluids in GRPA is explained 
from the absence of this exchange collision process. 

Nevertheless, atom exchanges between the normal and the super fluids should always exist 
in any kinetic description, in particular to guarantee the process of 
condensate formation. This is the case for the GRPA,
but provided that instability conditions are satisfied \cite{gap2}. 
For example, when the relative velocity between the two fluids 
exceeds the critical velocity given by the Landau criterion, 
the total screening phenomenon disappears and the binary collisions become 
again possible. 

\subsection{Experimental difficulties}
Both gap and total screening phenomena have been 
predicted to appear in a Raman transition process between two hyperfine 
levels of a $^{87}Rb$ gas, but only in the bulk 
case \cite{gap2}. 
In comparison to other methods like radio frequency (RF) or Bragg spectroscopy, 
the possibility of momentum transfer and the distinction 
between scattered and unscattered atoms  
enable these observations. 
However, an experimental realization is still 
not simple in the real case of a trap since the gas inhomogeneity,
combined with the short duration of the applied coupling potential, 
leads to additional broadenings of the spectral lines that prevent
the resolution of the gap and screening structure. In this context,
a RF spectroscopy 
would have probed the whole gas which includes thermal atoms 
of the outer and inner condensate regions. 
Therefore, the distinction between various theoretical approaches 
is extremely difficult  as long as  the transition amplitude  and  
the dispersion relation of thermal atoms have a strong spatial dependence.

\subsection{Setup proposals}

In this letter, we propose different methods to  
probe the atoms more efficiently than the RF spectroscopy: 
1) the Raman scattering is a two-photon process that 
offers also the possibility to transfer the momentum $\vc{q}$ to the 
scattered atoms and observe their resulting momentum distribution 
after expanding the gas; 2) a four photon scattering process, where 
two sets of two beams cross in the trap center, 
addresses selectively the homogeneous region of the gas (see Fig.\ref{f4}).

We apply these methods for the case of a finite temperature trapped Bose gas in the GRPA, 
in a bid to challenge the Bogoliubov approach. 
To this end, we propose two concrete experimental setups 
that overcome the difficulties associated with the trap:
1) The gap is observed from the four-photon process; 2) The total screening 
is determined in a Raman scattering.
Previous theoretical works \cite{condenson,gap2} argue in favor of the 
{\it conserving} GRPA. Nevertheless,  
a comparison with the {\it non conserving} 
Bogoliubov approximation is of relevance as long as the second 
branch of individual excitations has not been observed.



\section{Raman scattering}

\subsection{The GRPA approach}
\begin{figure}\label{vc}
\begin{center}
\resizebox{0.50\columnwidth}{!}{\includegraphics{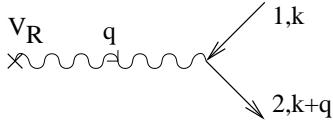}}
\end{center}
\caption{Diagrammatic representation of the scattering of an atom by an external 
potential. An 
atom of momentum $\vc{k}$ is scattered into a state of momentum $\vc{k+q}$ 
by means of an external interaction mediated by a virtual 
collective excitation of momentum $\vc{q}$.}
\label{f2}
\end{figure}

In a Raman transition,
we start from atoms initially in the hyperfine level $|1\rangle =
|F=1,m_F=-1\rangle$.  Each mode $\vc{k}$ is characterized by its initial population $N_\vc{0}$ 
and 
$N_{\vc{k}\not=0}=1/(\exp[\beta(\epsilon^{HF}_{1,\vc{k}}-\mu)]-1)$ and its initial
plane wave function $\psi_{1,\vc{k}} =
{\exp[i(\vc{k}.\vc{r}-\epsilon^{HF}_{1,\vc{k}} t)]}/{\sqrt{V}}$ with the inverse 
temperature $\beta=1/k_B T$ and the chemical potential $\mu=g(2n-n_\vc{0})$.
The application of a perturbation coupling potential $V_{\vc{q}}(\vc{r},t)=
V_R \exp[i(\vc{q}.\vc{r}-\omega t)]$ at $t \geq 0$ transfers a small fraction of them 
into the second level $|2\rangle =|F=2,m_F=1\rangle$ of internal frequency 
$\omega_0$. The determination of the second spinor 
component of the associated wavefunction 
$\psi_{2,\vc{k}}(\vc{r},t)$ of the mode $\vc{k}$
evolves according to the time-dependant Hartree-Fock equation \cite{gap2}:
\begin{eqnarray}
\label{p21}
\left[i\hbar{\partial_t}
+\frac{\hbar \nabla^2_\vc{r}}{2m}-\hbar\omega_0 -
g_{12}\sum_\vc{k'\not= \vc{0}} N_\vc{k'}
{|\psi_{1,\vc{k'}}|^2}\right] \psi_{2,\vc{k}}
= \nonumber \\
\left[V_{\vc{q}} +
g_{12}\sum_\vc{k'}  N_\vc{k'}
{\psi^{*}_{1,\vc{k'}}} \psi_{2,\vc{k'}}\right] 
\psi_{1,\vc{k}}
\end{eqnarray}
where we define the  
intercomponent coupling $g_{12}=4\pi \hbar^2 a_{12}/m$.
The solution is \cite{gap2}:
\begin{eqnarray}\label{psiR}
\psi_{2,\vc{k}}(\vc{r},t)=\!
\int_{-\infty}^\infty \!\!\!\!\!\!\!d\omega'
\frac{ \int_0^\infty \!dt' e^{i(\omega'+i0)(t'-t)}V_{\vc{q}}(\vc{r},t')\psi_{1,\vc{k}}(\vc{r},t)}
{2\pi i{\cal K}_{12}(\vc{q},\omega')(\hbar\omega'+i0-\hbar\omega_{\vc{k},\vc{q}})}
\end{eqnarray}
where $\hbar \omega_{\vc{k},\vc{q}}=\epsilon^{HF}_{2,\vc{k}+\vc{q}}-\epsilon^{HF}_{1,\vc{k}}$
and $\epsilon^{HF}_{2,\vc{k}+\vc{q}}=\hbar\omega_0+\epsilon_\vc{k+q}
+g_{12}(n- n_\vc{0}\delta_{\vc{k},\vc{0}})$ is the atom mean field energy in the second level 
without the exchange term. 
These formulae resemble the one obtained from the non interacting Bose gas except 
for the HF mean field terms  and 
the screening factor:
\begin{eqnarray}\label{K12}
{{\cal K}}_{12}(\vc{q},\omega)=1-\frac{g_{12}}{V}\sum_\vc{k}
\frac{N_{\vc{k}}}
{\hbar \omega +i0- \hbar \omega_\vc{k,q}} 
\end{eqnarray}
Eq.(\ref{psiR}) is interpreted in Fig.\ref{f2} in terms of 
propagators whose poles determine the resonance frequencies. 
One pole is associated to the individual transition between atoms: 
$\omega= \omega_{\vc{k},\vc{q}}$ and the other is the zero of the 
screening factor and corresponds to the collective excitations associated 
to the gas rotation in the spin space:
$\delta \omega =\omega-\omega_0 \sim [\epsilon_\vc{q}-(g-g_{12})n]/\hbar$ 
for $g_{12} \sim g$.
Total screening corresponds to the singularity
${{\cal K}}_{12}(\vc{q},\omega_{\vc{0},\vc{q}})
\rightarrow \infty$ and  
prevents any single condensed atom scattering
\cite{gap2}.

In a bulk gas, the transferred atom density for each mode is obtained from 
$n_{2,\vc{k}+\vc{q}}(t)=
|\psi^{(1)}_{2,\vc{k}}(\vc{r},t)|^2 N_\vc{k}$ 
so that we deduce the total atom density\cite{Stringari,gap2}:
\begin{eqnarray}
n_2=\! \sum_\vc{k} n_{2,\vc{k}}=\!\!
\int_{-\infty}^\infty
\!\!\!\!\!\! d\omega' \frac{4 \sin^2(\omega't/2)}{\hbar \pi \omega'^2}
|V_R|^2 \chi_{12}'' (\vc{q},\omega-\omega')
\end{eqnarray}
expressed in terms of the imaginary part of the intercomponent susceptibility 
function 
$\chi_{12} (\vc{q},\omega)=1/(g_{12} {{\cal K}}_{12}(\vc{q},\omega))$.

\subsection{The Bogoliubov approach} 

These results can be compared to the one obtained from the Bogoliubov 
{\it non conserving} approximation developed in \cite{Fetter,gap,gap2}  which is
valid only for a weakly depleted 
condensate. This approach implicitly assumes that the elementary excitations 
are the collective ones forming a basis of quantum orthogonal states for the  
description of the normal fluid. Consequently, this formalism predicts 
no gap and no screening. 
The creation-annihilation operators 
$c^\dagger_{i,\vc{k}}(t), c_{i,\vc{k}}(t)$ describing 
the various components in the momentum space evolve according 
to $c_{1,\vc{k}}(t)=e^{-i\mu t}(\sqrt{N_{\vc{0}}}\delta_{\vc{k},0}
+u_{+,\vc{k}} 
e^{-i\epsilon^B_{1,\vc{k}}t} 
b_{1,\vc{k}}+ 
u_{-,\vc{k}} e^{i\epsilon^B_{1,\vc{k}}t}
b^\dagger_{1,-\vc{k}})$ and $c_{2,\vc{k}}(t)=
e^{-i(\mu +\epsilon^B_{2,\vc{k}}) t}
c_{2,\vc{k}}$. In this expression, besides the collective 
excitation modes of phonon of energy $\epsilon_{1,\vc{k}}^B$, a 
second collective mode of rotation appears with energy $\epsilon_{2,\vc{k}}^B =\epsilon_\vc{k}+(g_{12}-g)n_\vc{0}$. $\mu=gn_\vc{0}$ is the chemical potential, 
$b_{1,\vc{k}}$ is the annihilation operator 
associated to the 
quasi-particle such 
that $\langle b^\dagger_{1,\vc{k}} b_{1,\vc{k}}\rangle= 
1/(\exp( \beta \epsilon^B_{1,\vc{k}})-1)$ and 
$u_{\pm,\vc{k}}=\pm
((\epsilon_\vc{k}+gn_{\vc{0}})/2\epsilon^B_{1,\vc{k}}\pm 
1/2)^{1/2}$. 
Reexpressing the intercomponent susceptibility 
\begin{eqnarray}\label{BP0}
\chi_{12}(\vc{q},\omega)=
\frac{i V}{\hbar} \int_0^\infty \!\!dt\
e^{i(\omega+i0)t}\langle [{\rho^{12}_{\vc{q}}}^\dagger(0),
\rho^{12}_{\vc{q}}(t)]\rangle
\end{eqnarray}
in terms of the autocorrelation function of the excitation operator 
$\rho^{\alpha \beta}_{\vc{q}}(t)=\sum_\vc{k}
c^\dagger_{\alpha,\vc{k}}(t)c^{}_{\beta,\vc{k+q}}(t)/V$, 
we calculate in the Bogoliubov approximation:  
\begin{eqnarray}\label{BP}
\chi^B_{12}(\vc{q},\omega) =\sum_{\pm,\vc{k}}
\frac{\delta_{\vc{k},\vc{0}} N_\vc{0}/2\pm u^2_{\pm,\vc{k}}/(\exp(\pm \beta \epsilon^B_{1,\vc{k}})-1)}
{V(\hbar \delta\omega
+i0 \pm\epsilon^B_{1,\vc{k}}-\epsilon^{B}_{2,\vc{k}\pm\vc{q}})}
\end{eqnarray}
In contrast to the GRPA, Eq.(\ref{BP}) describes a spin rotation transition 
of the condensed fraction, one transition involves the excitation transfer 
from a phonon mode into a rotation mode and another the excitation creation in the two modes 
simultaneously.

\subsection{Extension to the trap}

\begin{figure}
\resizebox{\columnwidth}{!}{
\includegraphics{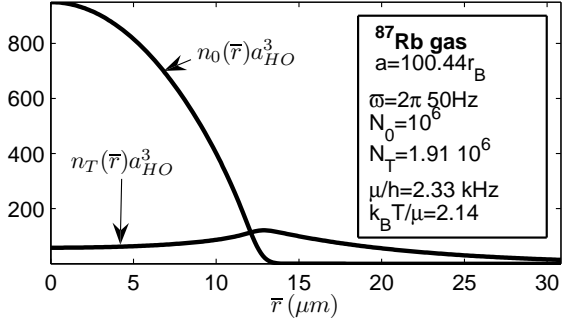}}
\caption{
Density profiles for the condensed and thermal clouds for   
typical values of gas parameters ($a_{H0}=\hbar/\sqrt{2m \overline{\omega}}$). 
$N_0=\int d^3\vc{r}\, n_0(\vc{r})$ and $N_T=\int d^3\vc{r}\, n_T(\vc{r})$ are the total 
number of condensed and thermal atom respectively.}
\label{f3}
\end{figure}

These formulae can be easily extended to the case of a harmonic trap 
$V_H(\vc{r})=\sum_i m\omega_i^2 r_i^2/2$ of frequency $\omega_i$ by considering 
the local density approximation (LDA) \cite{books}.
For a weakly inhomogeneous gas, the population 
in each mode becomes a local quantity $N_\vc{k} \rightarrow  N_\vc{k}(\vc{r})$.
By making this replacement, the thermal density 
$n_T(\vc{r})= \sum_{\vc{k}\not=0} N_\vc{k}(\vc{r})/V$, the energies
$\epsilon^{HF}_{i,\vc{k}}(\vc{r})$, $\epsilon^B_{i,\vc{k}}(\vc{r})$, 
the screening factor 
${\cal K}_{12}(\vc{r},\vc{q},\omega)$, the potential amplitude
$V_R(\vc{r})$ and $n_{2,\vc{k}}(\vc{r},t)$ become local 
quantities as well. 
The zero mode 
density $n_\vc{0}(\vc{r})=|\Psi_\vc{0}(\vc{r})|^2$ 
is determined from: 
\begin{eqnarray}\label{super2}
-\frac{\hbar^2 \nabla_\vc{r}^2\Psi_\vc{0}(\vc{r})}{2m\Psi_\vc{0}(\vc{r})}
+V_{H}(\vc{r})+g
(|\Psi_\vc{0}(\vc{r})|^2 +
2n_{T}(\vc{r}) )
=\mu
\end{eqnarray}  
while the non zero ones are determined from the 
semi-classical expression: 
\begin{eqnarray}\label{neqin}
N_{\vc{k}}(\vc{r})=
\frac{1}{\exp{[\beta(\epsilon^{HF}_{1,\vc{k}}(\vc{r})+V_{H}(\vc{r})-\mu)]}-1}
\end{eqnarray}
The set of Eqs.(\ref{super2},\ref{neqin}) is reduced to a one dimensional 
problem if we assume the ansatz $n_{\vc{0}}(\overline{r})$ where $\overline{r}
=\sqrt{2m V(\vc{r})}/\overline{\omega}$ and 
$\overline{\omega}=(\omega_x \omega_y \omega_z)^{1/3}$. 
This ansatz 
is exact for a spherical trap and is accurate in the Thomas-Fermi limit 
$\omega_i \ll gn(\vc{0})$. It leads to the
profiles in Fig.\ref{f3} for the condensed and normal fluids and shows
excellent agreements with both
experiments \cite{Aspect} 
and exact Monte-Carlo calculations \cite{Holzmann}
in the determination 
of the density profile of a trapped Bose condensed 
gas. 
These generalizations allow the determination of the 
transferred momentum distribution  
$N_{2,\vc{k}}(t)=
\int d^3\, \vc{r} \,n_{2,\vc{k}}(\vc{r},t) $
from which we deduce
the transferred thermal atom number $N_{2,T}(t)=\sum_\vc{k\not=\vc{q}}N_{2,\vc{k}}(t)$. 

\section{Experimental proposals}
\subsection{The gap experiment: four-photon process}
 
For $\vc{q}=0$ and $g_{12} \sim g$, the Raman spectrum 
becomes discrete in a homogeneous gas. The resonance frequencies 
correspond to a gap $\hbar \omega_{\vc{k},\vc{q}}(\vc{r})=-g n(\vc{r})$ 
associated to the exchange interaction energy for the single 
mode transition and to $(g_{12}-g)n(\vc{r})$ for 
the collective mode transition \cite{Levitov}. In comparison, 
if the condensed atom spectrum is quite similar, 
the thermal atom one displays differences 
in the Bogoliubov approximation. Since the energy difference 
$\epsilon_{2,\vc{k}}(\vc{r})-\epsilon^B_{1,\vc{k}}(\vc{r})$ is $\vc{k}$ 
dependant, no gap is observed and the oscillations are smoothed out 
leading to a continuous spectrum.  

\begin{figure}
\resizebox{\columnwidth}{!}{
\includegraphics{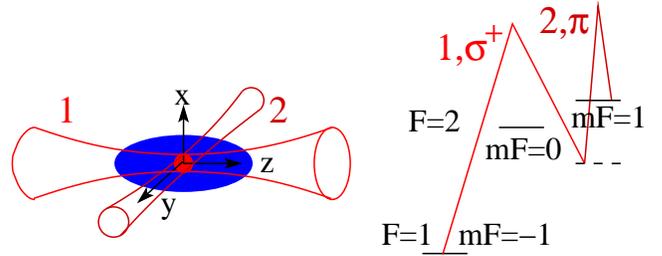}}
\caption{Selective four lasers interaction
with atoms within the trap center region.
Two lasers $\sigma^+$ polarized along the $z$ axis (1) drive the atoms through 
the intermediate states $m_F=0$ and $m_F=1$ while two others along 
the $y$ axis (2) drive the atoms within the sublevel $m_F=1$.} 
\label{f4}
\end{figure}

\begin{figure}
\resizebox{\columnwidth}{!}{
\includegraphics{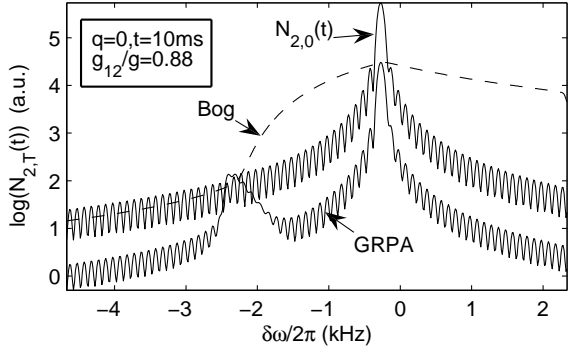}}
\caption{Transferred thermal and condensed atoms vs. the detuning in 
the GRPA and Bogoliubov approximation.} 
\label{f5}       
\end{figure}

In order to distinguish clearly between the discrete and the continuous 
spectra, the coupling potential acts specifically in the trap center in order 
to reduce inhomogeneous broadening.
This is realized by means of four beams (see Fig.\ref{f4})
\cite{Becker}: two  
gaussian astigmatic beams $\sigma^+$ polarized along the z axis 
of quantization with the intensity profile
$I_1(\vc{r})=I_{01} \exp(-2 r_x^2/w_1^2(r_z)-2r_y^2/w_2^2(r_z))$ and two others 
$\pi$ polarized along the y axis with 
$I_2(\vc{r})=I_{02} \exp(-2 r_x^2/w_3^2(r_y)-2r_z^2/w_4^2(r_y))$
where $w_i(s)=w_i (1+(s\lambda)^2/(\pi^2 w_i^4))^{1/2}$. The sum 
of their frequency differences corresponds to the transition frequency $\omega$. 
Provided that $\lambda \ll w_i$, we define an effective waist 
$\overline{w}$ such that: 
\begin{eqnarray} 
\frac{1}{{\overline{\omega}}\, {\overline{w}}^2}=
\frac{1}{\omega_x}(\frac{1}{w_1^2}+\frac{1}{w_3^2})=
\frac{1}{\omega_y w_2^2}=
\frac{1}{\omega_z w_4^2}
\end{eqnarray} 
In these conditions, the resulting potential 
$V_R(\vc{r})=V_{R0}\exp(-2\overline{r}^2/\overline{w}^2)$ 
is optimized 
for an atom transfer in the most homogeneous region with $\vc{q}=0$. 
To fix the idea, we choose $\lambda =843 nm$ and 
$\overline{w}= 7 \mu m$ which reduces to about $10^4$ the thermal atom 
effective number that can be specifically addressed.  
Transferring a small fraction of about 10\% and
for a detection resolution of about 100 atoms, we 
obtain a signal to noise ratio of about 10. 
A relative difference in the scattering lengths 
is also needed to observe the gap resonance and  is obtained 
from the application of an external magnetic field \cite{Kai}. 
These consideration leads to the spectra of Fig.\ref{f5}. 
Note the two orders of magnitude between the two peak intensities and the 
oscillatory behavior of period $1/t=100Hz$ associated to 
the finite time resolution. The finite size of the beam 
provides an additional negligible frequency uncertainty 
of about $\hbar/(\sqrt{m\beta}\overline{w})$ in the resolution.      

\subsection{The screening experiment: Raman scattering} 

The absence of 
Raman transition due to screening 
is observed in the scattered atom momentum distribution. 
For a long time, 
the transient effects in Eq.(\ref{psiR}) 
can be neglected leading 
to a constant transfer rate and, 
except for the fact that the external potential is screened, 
we recover the Fermi golden rule:
\begin{eqnarray}\label{FG}
\frac{N_{2,\vc{k+q}}(t)}{t}\stackrel{t \rightarrow \infty}{=}
\int d^3\vc{r}
\frac{2\pi V^2_R(\vc{r})n_{\vc{k}}(\vc{r})
\delta (\omega -\omega_{\vc{k},\vc{q}}(\vc{r}))}
{\hbar^2 |{\cal K}_{12}(\vc{r},\vc{q},\omega)|^2}
\end{eqnarray} 
Considering $g_{12} \sim g$, the transition
energy is position dependant causing inhomogeneous broadening:
$\hbar \delta \omega=k_z q_z/m+\epsilon_\vc{q}-gn(\vc{r})$. 
In the absence of screening, 
a resonance maximum appears for  $k_z=0$.
The screening factor strongly reduces the Raman
scattering and forbids it at this maximum i.e. 
$N_{2,k_x,k_y,q_z}(t)/t 
\stackrel{t \rightarrow \infty}{\rightarrow} 0$
thus avoiding the  
condensed atom transfer.
For 
simplicity, let $\omega_x=\omega_y$.
The atoms are transferred by means of
a Raman transition resulting from  
two gaussian symmetric laser beams such that their wavevector difference $\vc{q}$
is along the z axis and their frequency difference is the transition frequency $\omega$.
For small $q_z$, the angle between the beams is small and 
the Raman potential has the gaussian circular profile 
$V_R=V_{R0}\exp(-2(r_x^2+r_y^2)/w_5^2(r_z))$.
Once the atoms are transferred, the trap is switched off and after 
a time of flight, the density profile provides their momentum 
distribution. 

A negative 
detuning is chosen in order to scatter the thermal atoms with $k_z$ positive  
in the trap center region
and negative otherwise.
\begin{figure}
\resizebox{\columnwidth}{!}{
\includegraphics{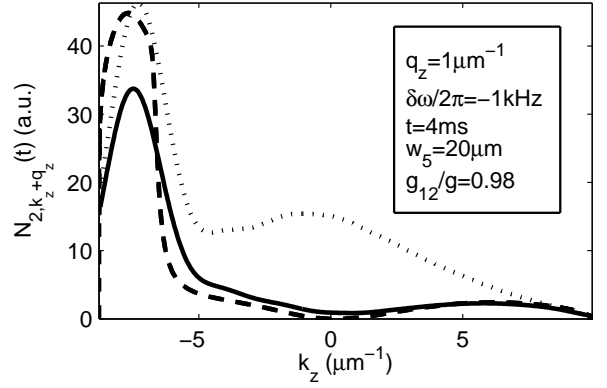}}
\caption{Thermal atom distribution $N_{2,k_z}=\sum_{k_x,k_y} N_{2,\vc{k}}$ versus $k_z$ 
in presence of screening using 
Eq.(\ref{FG}) (dashed line) and in absence (dotted line) and presence (full line) 
of screening taking into account finite interaction time corrections.}  
\label{f6}       
\end{figure}
The graphs in Fig.\ref{f6} illustrate well the total screening effect 
around $k_z=0$ for which the macroscopic 
wave function deforms its shape in order to attenuate 
locally the Raman potential, thus
preventing single atom scattering. 
The left 
part of the distribution ($k_z < -6\mu m^{-1}$) shows the thermal atoms coming from 
the outer condensate region. The choice of  
$q_z$ is such that
the LDA validity condition $q_z w_5 \gg 1$ is fulfilled 
but also such that, during the flight, the mean field energy 
does not affect much the momentum distribution. 
The interaction time must be much lower than the relaxation time 
associated with collisions $t \ll \tau \sim \sqrt{\beta  m/8\pi a^2 n_T(\vc{0})}$ 
to avoid the equilibrium relaxation of the momentum distribution.  
Its finite value creates an energy uncertainty 
that alters the 
validity of  Eq.(\ref{FG}) by not suppressing totally 
atom scattering at $k_z=0$. Also, this time must be adequately chosen 
to suppress the condensed fraction due to the Rabi flopping associated to the 
collective mode: 
$N_{2,\vc{q}}(t)=2\sin^2[(\hbar \delta \omega -\epsilon_\vc{q})t/2]N_{20,\vc{q}}$ 
where $N_{20,\vc{q}}\simeq 2\int d^3\vc{r}\, n_{\vc{0}}(\vc{r})
[V_R(\vc{r})/(\hbar \delta \omega -\epsilon_\vc{q})]^2=0.19 N_{2,T}$ for the case of Fig.\ref{f6}.

\subsection{Phenomenological approach}

Although theoretical statements argue in favor of GRPA, we cannot 
exclude that  none of the two approximations reproduces correctly 
the physical observation. In such a case, we can use a phenomenological 
approach assuming a transition process from an excitation of unknown 
energy $\epsilon_{1,\vc{k}}^X$ to an excitation of energy 
$\epsilon_{2,\vc{k+q}}^X$ and a process of creation of two excitations of energy 
 $\epsilon_{1,\vc{-k}}^X$ and $\epsilon_{2,\vc{k+q}}^X$. Using a 
four photon process interacting in the uniform region of the gas, 
the fraction of scattered atoms is then written 
under the form analog to Eq.(\ref{FG}): 
\begin{eqnarray}\label{phen}
\frac{N_{2,\vc{k+q_z}}}{t}\stackrel{t \rightarrow \infty}{\simeq}
\sum_\pm A_\pm (\vc{q}_z, \vc{k}) 
\delta(\omega \pm \epsilon_{1,\vc{\pm k}}^X +\epsilon_{2,\vc{k+q}}^X)
\end{eqnarray}
where $A_\pm (\vc{q}_z, \vc{k})$ represent the associated amplitude 
for such transition processes. Experimentally, the imaging in two dimensions 
allows only the determination of $F(k_x,k_z)=\int_\infty^\infty dk_y N_{2,\vc{k+q_z}}$. Thus 
the quantity (\ref{phen}) is determined from the Abel's transformation:
\begin{eqnarray}
N_{2,\vc{k+q_z}}=-\frac{1}{\pi}\int_{\sqrt{k_x^2+k_y^2}}^\infty
\frac{d F(y,k_z)}{dy}\,\frac{dy}{\sqrt{y^2-(k_x^2+k_y^2)}}
\end{eqnarray} 
By varying the parameters $q_z$ and $\omega$, the resonance positions in the 
$\vc{k}$ space allow 
to reconstruct the dispersion relations $\epsilon_{1,\vc{k}}^X$ and 
$\epsilon_{2,\vc{k}}^X$ for the excitations.

\section{Conclusions}

We explored the many body properties of a trapped Bose gas 
that can be extracted 
from a two-level hyperfine transition in the GRPA and Bogoliubov 
approximation.
The calculated spectra not only show the existence of a
second branch of excitation but also the total screening of the 
external potential
which prevents single condensed atom transitions.
If the external potential originates from the presence of 
a thermal atom, this total screening prevents 
the binary collision between that thermal atom and any condensed 
one. In this scenario, the metastability of the
relative motion between the normal and super fluids is explained 
by the absence of this exchange collision process \cite{condenson}.
The experimental observation of these phenomena will improve our understanding 
of the exact nature of the elementary excitations and of the origin of metastable 
motions in superfluids.

PN gratefully
acknowledges support 
from the Belgian FWO project G.0115.06, from the 
Junior fellowship F/05/011 of the KUL research council,  
and from the German AvH foundation. KB 
thanks EPSRC for financial support in grant 
EP/E036473/1.

\end{document}